\documentstyle[12pt,psfig]{article}
\begin{document}
\bibliographystyle{unsrt}
\pagenumbering{roman}
\baselineskip=.3in
\pagestyle{plain}
\def\be{\begin{equation}} \def\ee{\end{equation}}
\def\bfi{\begin{figure}[p]} \def\efi{\end{figure}} \smallskip
\def\la{\langle} 
\def\ra{\rangle} 

\pagenumbering{arabic}

\title{ {\Large \bf  The effects of large scales on the inertial range in high-Reynolds-number turbulence} }  
\author{Katepalli R. Sreenivasan, Brindesh Dhruva, and Inigo San Gil\\
Mason Laboratory, Yale University, New Haven, CT 06520-8286, USA\\
e-mail: k.sreenivasan@yale.edu}

\date{  }

\maketitle
\vskip 1cm

\abstract
{The effects of removing large scales external to the inertial range on the properties of scales within the inertial range are studied in a
high-Reynolds-number turbulent flow. Structure functions of
both even  
and odd orders are strongly affected across the entire inertial range, but  
odd-order moments are affected to a greater degree. In particular, the
skewness of  
velocity increments shows a significant reduction whereas the flatness
changes comparatively little. The reduction in skewness is counterbalanced essentially by the interaction between the  
small-scale energy and the large-scale rate of strain. The implications of these results for the conventional cascade picture are examined briefly.}

\bigskip
PACS Numbers: 47.27.Ak, 47.27.Jv

\bigskip
\baselineskip 0.25truein
\section {Introduction}

The K\'arm\'an-Howarth equation$^1$ is an exact dynamical equation for the
second-order  
correlation function in isotropic turbulence. The equation is easily
expressed in  
terms of the longitudinal structure function $S_2(r,t)$, where $S_2 \equiv \la \Delta u_r^2 \ra \equiv \la
[u(x+r) -  
u(x)]^2 \ra$, $u(x)$ is the turbulent velocity component in the direction
$x$, and  
$r$ is the separation distance in that same direction. Assuming that the
time  
variation of $S_2$ is slow, Kolmogorov$^2$ reduced the K\'arm\'an-Howarth
equation  
to an equation for the third-order structure function. This equation, given by
\be
S_3(r) \equiv \la [u(x+r) - u(x)]^3 \ra = -\frac{4}{5} \la \varepsilon \ra
r + 6  
\nu \frac{\partial S_2(r)}{\partial r},
\ee
has been rederived under less restrictive circumstances$^{3,4}$, and works
well in  
shear flows as long as $r$ is much smaller than a characteristic large scale, $L$. Here $\la \varepsilon \ra$ is the average
value of  
the energy dissipation rate. Figure 1 is a typical comparison between the
measured  
third-order structure function and the right hand side of Eq.~(1). In
particular,  
there is a range of scales over which the viscous term is negligibly small,
and we  
have
\be 
S_3(r) = -\frac{4}{5} \la \varepsilon \ra r. 
\ee
Indeed, it is conventional to define the inertial range as the range of
scales for  
which this equation is valid.

The usual interpretation of Eq.~(2) is that, on the average, it
prescribes a unidirectional down-scale energy flux.$^{5,3}$ It is
consequently thought that the statistical physics appropriate to the
inertial range has a strongly non-equilibrium character.$^6$ On the other
hand, consider the following argument. If the inertial range dynamics is
not influenced by viscosity (as apparent in the  
previous paragraph), it is reasonable to suppose that the dynamics there is
governed essentially by the Euler equations. The Euler equations possess  
time-reversal symmetry, $t \rightarrow -t$, $\bf u \rightarrow  \bf -u$ (t =  
time, {$\bf u$} = vector velocity), which could therefore be interpreted to
mean that the energy  
flow in the inertial range is essentially as much up-scale as down-scale. The
proper meaning of Eq.~(2) would then be that the energy flux it prescribes
is a small difference between equally large  
energy flows in the up-scale and down-scale directions. The finiteness of
$S_3$ may then mean that there exists no more than a ``small leakage" between two  
opposing and equally strong fluxes.$^7$ If this is true, the situation calls for a
simpler type of statistical physics. 

Prompted by such qualitative considerations, we studied, some nine years ago,$^8$ the source of energy flux in the Kolmogorov equation. A result of
particular interest is the experimental finding that $S_3$ became small
when large scales in the fluctuating velocity were filtered out.
This result was thought to imply that the Kolmogorov result, Eq.~(2), was
more a reflection on the source of the  
energy flux (large scales) than a statement about unidirectional cascades.
We present here an up-date of the experimental part of that (unpublished) study. We use  
velocity signals acquired in a very high-Reynolds-number atmospheric
surface layer,  
described below, to analyse the effect of removing large scales on the  
statistics of $\Delta u_r$, in particular the effects on its third moment.
The  
principal conclusion is that, while the removal of large scales affects both  
even-order and odd-order moments of velocity increments in the inertial
range, the  
odd moments are much more vulnerable. The results are interpreted briefly
in the  
light of the issues raised above, though it is fair to warn the reader that we
shall not be able to resolve the issue satisfactorily.
Section II is a brief commentary of the experimental data, while Sec.~III
presents the basic experimental results. The chief conclusions are
discussed in Sec.~IV.

\section{The data and the filter characteristics}

The filtering experiments to be described below have been made for a number
of flows  
such as a pipe flow (diameter Reynolds number $\approx$ 230,000), boundary layer  
turbulence (boundary layer thickness Reynolds number $\approx$ 20,000),
atmospheric  
turbulence at a height of 6 m above the ground (Taylor microscale
Reynolds  
number $R_{\lambda} \approx$ 2,000), and atmospheric turbulence at a height
of 35 m  
above the ground ($R_{\lambda} \approx$ 10,000).$^9$ The results are
entirely  
consistent with each other, but are more revealing when the Reynolds number
is high  
and the scale separation large, so we present data only for the last case.
Tests  
were made on several sets of data, but the results are presented for brevity for
only one.

The atmospheric data were acquired by a hot-wire probe mounted at a height
of 35 m  
above the ground on the meteorological tower at the Brookhaven National
Laboratory.  
The wind speed and direction were independently monitored by a vane
anemometer  
mounted close to the tower. During data acquisition, the mean wind was
essentially  
steady in both magnitude and direction. This was ensured, {\it a posteriori},
by using  
only those sets of data for which the averages computed over different
segments of  
the measurement interval were steady. The conditions of measurement thus
approached  
`controlled' circumstances.

The hotwire, about 0.7 mm in length and 5 $\mu$m in diameter, was operated
in a  
constant temperature mode. Its frequency response was compensated to be
flat up to  
20 kHz. The voltage from the anemometer was low-pass filtered and
digitized. The  
low-pass cut-off was half the sampling frequency, $f_s$, which was set at
5,000 Hz.  
This frequency was high enough to include essentially all the finest-scale
turbulent  
fluctuations in the flow. The voltage was constantly monitored on an
oscilloscope  
to ensure that it did not exceed the digitizer limits. The voltage from the  
anemometer was converted to wind velocity through the standard calibration  
procedure, which included calibrating the hotwire just prior to mounting and  
checking it immediately after dismounting. Taylor's frozen flow assumption
was  
employed to convert the time variable into the spatial distance. 

The following characteristics pertain to the present data: The mean wind
speed  
$\overline U$ $\approx$ 7.6  ms$^{-1}$, root-mean-square fluctuation velocity
$u^\prime \approx$  
1.36 ms$^{-1}$, the mean energy dissipation rate $\langle \varepsilon
\rangle \approx 3  
\times 10^{-2}$ m$^2$s$^{-3}$, the Kolmogorov scale $\eta \approx$ 0.57 mm, the
Taylor  
microscale in the streamwise direction $\lambda \approx 11.4$ cm. The Taylor
microscale  
Reynolds number $u^{\prime} \lambda/\nu \approx 10,340$. We use the height from
the  
ground, $H$, as a measure of the large scale. If we assume that the classical
boundary  
layer arguments are valid, a characteristic scale for the large-scale
transport is  
of the order $0.4 \times 35$ m, where 0.4 is the `accepted' numerical value of the
K\'arm\'an  
constant.

To obtain the filtered signal we used a Butterworth
filter  
with excellent cut-off characteristics and no phase shift. The filtering operation was done by convolution in real space. The cut-off scale $r_f$ and the cut-off frequency $f_o$ are related by $r_f = U/f_o$. After filtering in the forward
direction,  
the filtered sequence is reversed and run back through the filter. The
resulting  
sequence has no phase distortion and double the filter order.  We have used
filters of order 4 and 10 (i.e., of order 2 and 5, respectively, but with
two  
passes), mostly the latter. The filter characteristics are demonstrated in
Figs.~2  
to 4. (A different filtering scheme was employed in Ref.~[8] and the nature of the results was identical.) Figure 2 shows the energy spectral density of streamwise velocity  
fluctuations. The Kolmogorov scaling of the spectral density for the
unfiltered signal  
extends roughly over three decades of scales. The spectrum for the filtered
signal,  
with high-pass setting at 2 Hz is also shown for the tenth-order filter. The  
attenuation in Fourier space exceeds 40 dB per octave. The ratio of the power
spectral  
density of the filtered signal to that of the unfiltered signal is shown in
Fig.~3  
for fourth-order and tenth-order filters. For the latter, the filter effects
vanish by  
within a factor of 2 of the cut-off setting.  

The filter produces minimal phase distortion, as 
demonstrated below. Let $u$ be the unfiltered velocity, and $u^>$ and $u^<$ be the  
high-pass (small-scale) and low-pass (coarse-scale) parts of $u$, both obtained by independent  
filtering operations with the filter set at $f_o$ Hz. Define $u_*^< = u -
u^>$.  If  
the filter does not produce any phase distortion, we should find $u^< = u_*^<$. Figure 4 shows this is indeed so.

\section{Experimental results}

\subsection{Probability density functions}

It is known$^{10-12}$ that the probability density function (pdf) of the
velocity  
increment, $\Delta u_r$, continually changes its character as the separation  
distance $r$ is varied: it is roughly square-root exponential for $r$ in the  
dissipative range and Gaussian when $r$ is comparable to the large  
scale. The pdfs are shown in Fig.~5 (a).   
Figures~5 (b), (c) and (d) show similar data for high-pass filtered signals, each for
a different  
filter setting.  The spatial cut-off scale $r_f$ is given in the caption. The removal of the large scales changes the
pdfs in two  
respects: (1) the  
shape is affected significantly for separation distances $r$ smaller than the
smallest  
scale removed by filtering, the latter being of the order $r_f$; (2) the asymptotic form of the pdf, for $r > r_f$, does not differ much from that at $r = r_f$, and is far from Gaussian. These effects can be quantified by fitting a
stretched  
exponential, $p_{\Delta u_r} = p(0) exp[-c|\Delta u_r|^{m(r)}]$, to the pdfs
of  
filtered signals, and comparing the stretching exponent $m(r)$ for various
filter  
settings. This is done in Fig.~6. The exponent
$m$ is  
slightly different for positive and negative sides of the distribution (shown by full and dashed lines, respectively), but
these  
differences do not mask the major effect: the exponent is noticeably
smaller for  
filtered signals than for the unfiltered signal, and the effect
occurs at  
scales significantly smaller than the smallest filtered scale.

\subsection{The third-order structure function}

Figure 7 is a summary of the effects of the  
filtering operation on the third-order structure function. The ordinate is  
normalized by $r \la \varepsilon \ra$, and the separation distance is
normalized by  
the Kolmogorov scale $\eta$ for the unfiltered data. First, a remark is in order on
the function $K \equiv \langle\Delta u_r^{3}\rangle/r \langle \varepsilon
\rangle$  
for the unfiltered data: the inertial range, defined as the region where
$K$ is  
flat and equal to 4/5, ranges roughly between $100\eta$ and $2500\eta$. (Note
that  
the spectral density shown in Fig.~1 appears to possess a larger scaling range,
but it is  
more prudent to base this estimate on the basis of Eq.~(2).)

The high-pass filtered data are obtained by removing all scales above 30 m (frquencies below about 0.25 Hz), 3.8 m ($\sim$ 2 Hz) and  
1.5 m ($\sim$ 5 Hz). The arrows mark the values of $r_f/\eta$. For the 3.8 m case, for example, $\la \Delta u_r^3 \ra$ is smaller
by an  
order of magnitude for $r/r_f = 0.5$, by a factor 2 for $r/r_f = 10$, and
by about  
10\% for $r/r_f = 100$. Inferring roughly from Fig.~2 that the artifacts of the
filtering operation do not extend to scales smaller than about $r_f/2$, it is clear
that the removal of  
large scales beyond the upper edge of the inertial range ($\sim$ 1.7 m) has a strong effect  
throughout the inertial range. However, the dissipation scales are
not  
affected significantly, so $\la \varepsilon \ra$ and $\eta$ remain essentially the same as for the unfiltered data; we have therefore simply used the unfiltered
values for  
normalization. The dissipation scales just begin to feel the effect of
filtering for  
$r_f$ = 1.5 m, but this cut-off begins to encroach directly on the inertial
range itself.  For most of the results below, we shall use $r_f$ = 3.8 m ($\sim f_o$ = 2 Hz), unless stated otherwise.

A naive interpretation of Fig.~2 is that the large scales contribute significantly to $\la \Delta u_r^3 \ra$ in the inertial range. A closer look demands a more  
complex interpretation.  By splitting, as before, the velocity signal $u$ into
the fine-scale (high-frequency) and  
coarse-scale (low-frequency) velocity components $u^>$ and $u^<$, separated by the cut-off scale $r_f$ or the cut-off frequency
$f_o$,  
and noting that $u = u^> + u^<$ and  $\Delta u_r = \Delta u_r^> + \Delta
u_r^<$,  
the third-order structure function can be written as
\begin{equation}
\langle \Delta u_r^3 \rangle = \langle (\Delta u_r^>)^3 \rangle + 3 \langle
(\Delta  
u_r^>)^2 \Delta u_r^< \rangle + 3 \langle (\Delta u_r^<)^2 \Delta u_r^>
\rangle +  
\langle (\Delta u_r^<)^3 \rangle.
\end{equation}
In the above equation, it is not hard to argue that $\la (\Delta u_r^<)^3 \ra$ and 3$\la  
(\Delta u_r^<)^2 \Delta u_r^> \ra$ should
both be  
small for very small $r$, and that the term $\la (\Delta u_r^>)^3 \ra$
should approach $\la \Delta  
u_r^3 \ra$ itself as $r \rightarrow 0$. For intermediate values of $r$, it
is not  
easy to estimate the correct orders of magnitude of these terms.
Experimental data  
(see Fig.~8) show that the largest term is
$\la  
(\Delta u_r^>)^3 \ra$ in the lower part of the inertial range and 3$\la
(\Delta  
u_r^>)^2 \Delta u_r^< \ra$ in the upper part, until they are both
overtaken, as $r  
\rightarrow r_f$, by the exclusively large-scale part $\la (\Delta u_r^<)^3 \ra$.   

The dominance of the term 3$\la (\Delta u_r^>)^2 \Delta u_r^< \ra$ suggests
that a  
significant part of the third-order structure function comes from the straining of the
small-scale  
kinetic energy, $(\Delta u_r^>)^2$, by the large-scale feature
$\Delta u_r^<$. This suggests the existence of a coupling between scales that are significantly separated. The effect is illustrated in a different manner in Fig.~9 where the second  
order moment of $\Delta u_r$, conditioned on the velocity gradient of the cut-off scale, $d u_r^</d r_f$, is plotted
against the latter quantity; the conditioning variable is the gradient of the largest scale rejected by the filtering operation. The variance may be expected to be independent of the conditioning variable if the energy transfer across scales were purely a
 local phenomenon---in contrast to the measured effect.

\subsection{Odd versus even moments}

The second and fourth order structure functions for the unfiltered and
filtered  
data are shown in Fig.~10. Filtering has
some  
effects on even-order moments as well; the larger the moment order,  
the larger this influence. However, these effects are much weaker than that for
the third-order  
moment (and for other odd-order moments as well, though not shown here). Two
specific statements can be made. First, for the second-order structure
function, the filter is felt only up  
to scales that are at best an order of magnitude smaller than $r_f$, but the effect
on the  
third-order extends to scales that are smaller by one further order of
magnitude.   
Second, the effects on the third-order are larger than those on the next-higher even moment, namely the fourth.

A better appreciation for differences between odd and even moments can be
gained by  
examining normalized moments. Figure~11 shows
the  
skewness of the velocity increment for the three filter settings used
above,  
as well as for the unfiltered data.  A comment first on the unfiltered signal itself: the fact that the skewness of the  
unfiltered signal is not constant in the inertial range shows that the  
$r$-dependence of $\la \Delta u_r^2 \ra$ has a power-law exponent that is
higher  
than 2/3. While a rough estimate of this ``anomalous correction" for
second-order  
structure function can be obtained from this graph, it has been argued  
elsewhere$^{9}$ that a proper estimate needs greater care. This is not our
main  
point, so we shall make no further reference to it here. Note that the
effect of  
removing scales above 30 m (below 0.25 Hz) already penetrates
the inertial range; removing scales above 3.8 m (below 2 Hz) and 1.5 m (5 Hz) significantly reduces the skewness of scales
everywhere in the  
inertial range. For instance, the skewness for the 2 Hz case is reduced by
a factor  
of 2 for scales one order of magnitude smaller than $r_f$.  On
the other  
hand, the flatness of $\Delta u_r$ shows hardly any effect for scales
smaller than  
$r_f/2$ (Fig.~12).

\section{Discussion and conclusion}

A principal conclusion of this work is that the removal of large scales
significantly affects the statistics well into the inertial range. In terms
of the
normalized moments of velocity increments, the skewness---as are other odd
moments---is affected much more strongly  
than the flatness---or other even moments. Indeed, if the large scales can
be thought of as related, loosely, to the ``size of the system", it appears that the system-size effects are felt more strongly on odd moments than on
even moments of velocity increments. 

It is worth asking if this effect is special to shear flows, because all the flows we have explored are shear flows. But we have some experience in assessing the shear effects$^9$ on scaling, and this experience suggests that the present conclusion is not
 restricted. It would, however, be worth repeating these calculations for isotropic turbulence. We have not done so because all the isotropic data available to us---experimentally as well as by simulations---possess limited scale separation, which makes i
t very difficult to extract meaningful results.  

It must be emphasized that the present results do not suggest that the skewness of the velocity increment
resides in the large $r$. Rather, what they suggest is that this asymmetry
of the pdf is related to the straining of the energy in the inertial range
by large scales. The asymmetry is the subject of another study in its own
right,$^{13}$ but it should be pointed out that it is associated with all
magnitudes of $\Delta u_r$. To see this, consider the difference between
the two sides of the pdf of $z \equiv \Delta u_r/(r
\la \varepsilon \ra)^{1/3}$. The quantity $z^3 \times [p(z) - p(-z)]$, when
integrated over $0 < z < \infty$, yields the function, $K \equiv \la \Delta u_r^3 \ra /r \la \varepsilon \ra$.
This integrand $z^3 \times [p(z) - p(-z)]$ is plotted against $z$ in
Fig.~13. The figure shows, for these specific conditions, that most of the contribution to $K$ comes largely from $z $ lying between about 2 and 10. The small values of $z$ lying between 0 and
2 make the opposite contribution to $K$. The filtering
operation essentially affects the integrand everywhere.

What do the present results imply for the energy cascade and the nature of
non-equilibrium (or otherwise) prevailing in the inertial range? They show that there are significant nonlocal effects, and that the energy cascade, even as an average concept, is not driven entirely by local effects. While cascade models do have pedagogi
cal value, they seem to miss this element. However, interpretations on the non-equilibrium nature of inertial-range dynamics
are difficult because the information at hand is very limited.
While it is unclear that the time reversal
symmetry of the Euler equations guarantees the near-equality of fluxes in
both directions (for instance, Euler equations can in principle
produce energy flux by an inviscid mechanism$^{14}$), it is equally
unclear that Eq.~(2) in itself demands a strong non-equilibrium
situation in the inertial range.

As a contrast, it is worth remembering that the subtle manner in which irreversibility appears even in the simpler case of the kinetic theory of
gases took a long time to unfold. One should therefore not feel too pessimistic about the inconclusiveness of the present situation. 

\vspace{0.3in}

\noindent {\bf Acknowledgements}\\

The work arose initially from a collaboration with A.A. Migdal and V.
Yakhot  
(theory) and with P. Kailasnath and L. Zubair (experiment). We express our
sincere  
thanks to them. Since the time of writing the unpublished
paper$^8$  
some nine years ago, we have discussed various aspects with A.J. Chorin, U. Frisch, R.H. Kraichnan, M. Nelkin, G. Parisi, I.
Procaccia, and  
E.D. Siggia. We are grateful to them for their generous comments.

\newpage

\noindent{\bf References}

\vspace{0.2in}

$^1$K\'arm\'an, T., von and L. Howarth, ``On the statistical theory of
isotropic  
turbulence", {\it Proc. Roy. Soc.} {\bf A164}, 192-215 (1938).

$^2$A.N. Kolmogorov, ``Energy dissipation in isotropic turbulence", {\it
Dokl.  
Akad. Nauk. SSSR}, {\bf 32}, 19-21 (1941).

$^3$U. Frisch, {\it Turbulence: The Legacy of A.N. Kolmogorov,} Cambridge  
University Press, Cambridge, 1995.

$^4$R. Hill, ``Applicability of Kolmogorov's and Monin's equations of turbulence", {\it J. Fluid Mech.}, {\bf 353}, 67-81 (1997).

$^5$A.S. Monin and A.M. Yaglom, ``{\it Statistical Fluid Mechanics: Mechanics of  
Turbulence}", vol. 2, M.I.T. Press, Cambridge, MA, 1975.

$^6$R.H. Kraichnan and S. Chen, ``Is there a statistical mechanics of turbulence?", {\it Physica D}, {\bf 37}, 160-172 (1989).

$^7$A. Chorin, {\it Vorticity and Turbulence}, Springer-Verlag, New York, 1994.

$^8$P. Kailasnath, A.A. Migdal, K.R. Sreenivasan, V. Yakhot and L. Zubair,
``The  
4/5-ths Kolmogorov law and the odd-order moments of velocity differences in  
Turbulence", unpublished report, Yale University (1991), pp. 18. 

$^{9}$K.R. Sreenivasan and B. Dhruva, ``Is there scaling in
high-Reynolds-number  
turbulence?", {\it Prog. Theo. Phys. Suppl. 130}, 103-120 (1998).

$^{10}$B. Casting, Y. Gagne and E.J. Hopfinger, ``Velocity probability density functions of high-Reynolds-number turbulence," Physica, {\bf 46D},  
177-200 (1990).

$^{11}$P. Kailasnath, K.R. Sreenivasan and G. Stolovitzky, ``The probability
density of  
velocity increments in turbulent flows", {\it Phys. Rev. Lett.}, {\bf 68},  
2766-2780 (1992).

$^{12}$P. Tabeling, G. Zocchi, F. Belin, J. Maurer and H. Williame,
``Probability  
density functions, skewness and flatness in large Reynolds number
turbulence", {\it  
Phys. Rev. E.} {\bf 53}, 1613-1621 (1996).

$^{13}$ S.I. Vainshtein and K.R. Sreenivasan, ``Kolmogorov's 4/5-ths law and intermittency in turbulence", {\it Phys. Rev. Lett.}, {\bf 73}, 3085-3089 (1994).

$^{14}$L. Onsager, ``Statistical hydrodynamics", {\it Nuovo Cimento}, {\bf 6} (suppl.) 279-287 (1949). 

\newpage
\begin{figure}[ht]
\begin{center}
\parbox{5.5in}{
\psfig{file=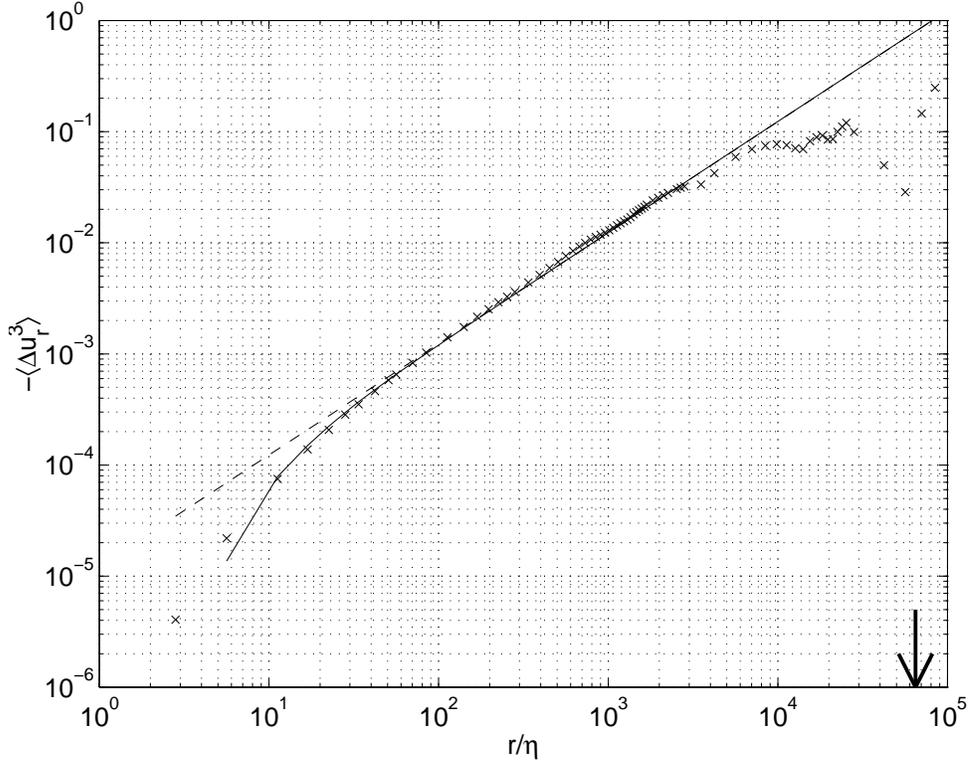,width=5.5in,rwidth=5.5in}
}
\caption{\small{The (negative value of the) third-order structure function, $S_3$, plotted against the separation  
distance, $r$, normalized by the Kolmogorov scale $\eta$. The measured data are compared with Eq. (1), full line. The  
agreement is quite good up to $r = H/20$. The largest
disagreement, seen at the very small scales, is due to the difficulty of resolving those scales well. The dashed line, which differs from Eq.~(1) only below $r/\eta$ of about 50, is Eq.~(2).  The data are for the atmospheric surface layer at the height $H
 = 35$ m above the ground, and all averages are over time; details of measurements are given in Sec. II. Here, $H$ is treated as the representative large scale, and marked by an arrow on the abscissae.}}
\label{fig:third_order_stfc}
\end{center}
\end{figure}

\newpage
\begin{figure}[ht]
\begin{center}
\parbox{5.5in}{
\psfig{file=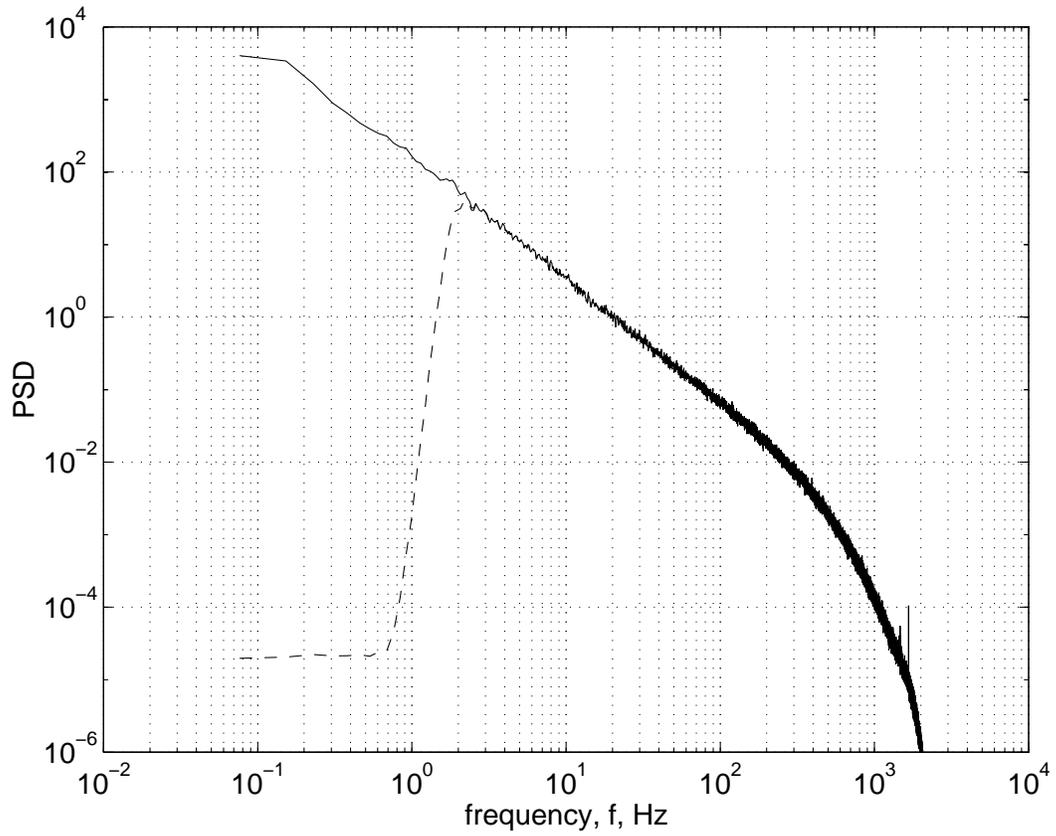,width=5.5in,rwidth=5.5in}
}
\caption{\small{The full line is the power spectral density (PSD) of the
longitudinal (streamwise) velocity fluctuation. The dashed line is 
the effect of a high-pass  
filter that is set to $f_o$ = 2 Hz. The filter is sharp and does not
produce oscillations.}}
\label{fig:spectrum_filter_unfilter}
\end{center}
\end{figure}

\newpage
\begin{figure}[ht]
\begin{center}
\parbox{5.5in}{
\psfig{file=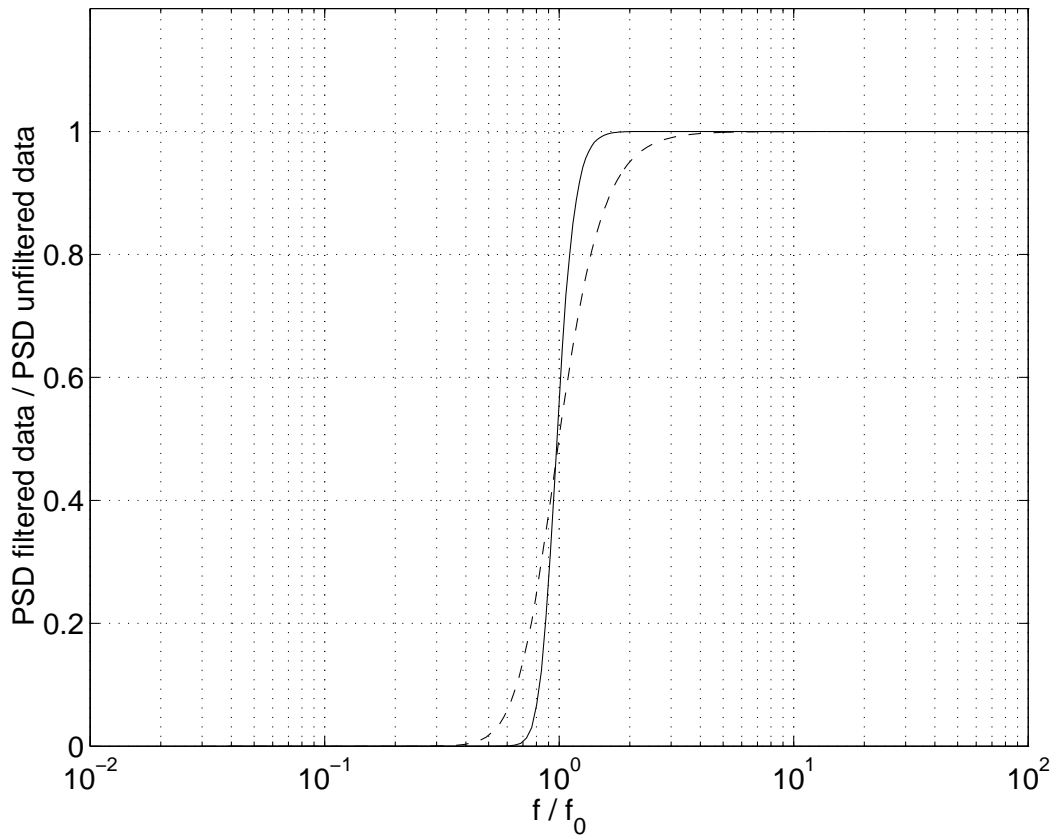,width=5.5in,rwidth=5.5in}
}
\caption{\small{The ratio of the filtered to the unfiltered power spectral
densities for  
the fourth and tenth order filters. We have experimented with both filters but used  
the tenth-order filter for the data presented in the rest of the text. For
this filter, the filter effects essentially disappear for frequencies
smaller than about $f_o/2$.}}
\label{fig:ratio_spectrum_filter_unfilter}
\end{center}
\end{figure}

\newpage
\begin{figure}[ht]
\begin{center}
\parbox{5.5in}{
\psfig{file=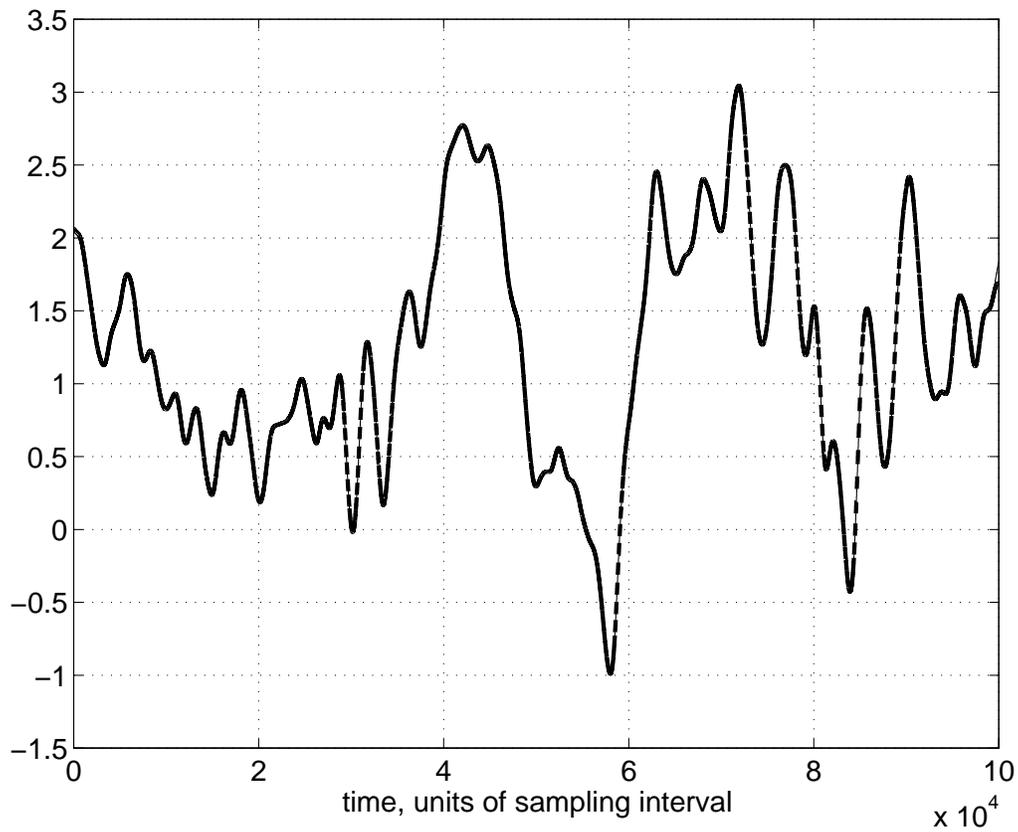,width=5.5in,rwidth=5.5in}
}
\caption{\small{For a segment of the data, this figure compares the low-pass filtered part, $u^<$, with $f_o$ set at 2 Hz (full line), with $u_*^< \equiv u - u^>$ where $u^>$ is the
high-pass filtered part obtained by an independent filtering operation, also performed with $f_o$ set to 2 Hz (heavy dashed line). The agreement between the two is excellent, as should be expected if the  
filtering operation produced no significant phase distortion. }}
\label{fig:signal_filter_unfilter}
\end{center}
\end{figure}

\newpage
\begin{figure}[ht]
\begin{center}
\parbox{5.5in}{
\psfig{file=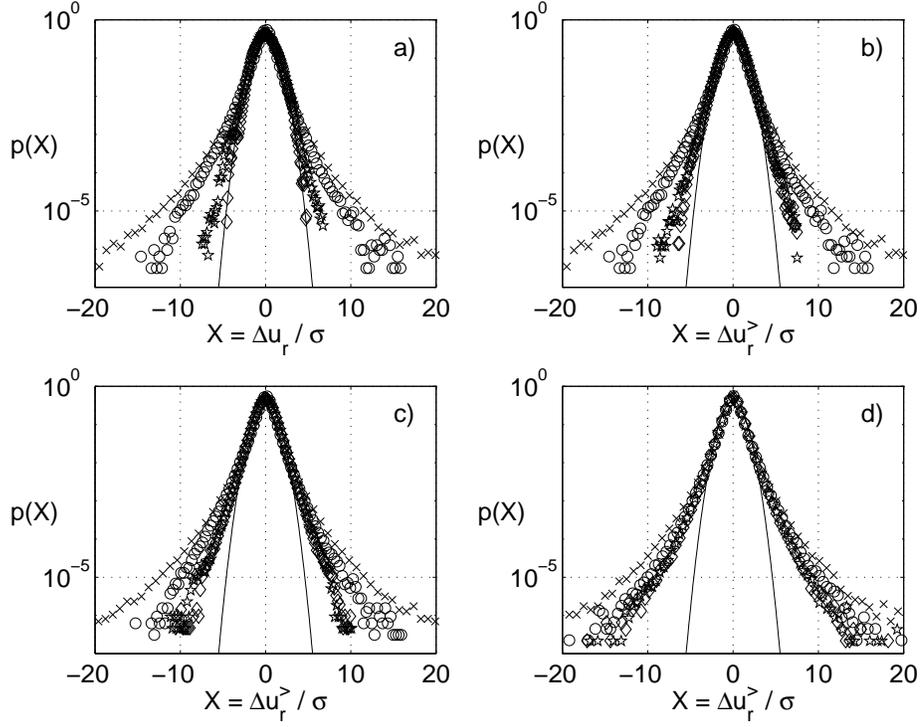,width=5.5in,rwidth=5.5in}
}
\caption{\small{The effect of filtering on the probability
density  
functions of $\Delta u_r$. In each case, the pdf is normalized by its own standard deviation, generically denoted by $\sigma$. The pdfs correspond to the following separation distances, $r$. $\times$: $r/H \approx 2.2 \times 10^{-4}$, $r/\eta \approx 13$;
 $\circ$: $r/H \approx 3
\times  
10^{-3}$, $r/\eta \approx 186$; $\star$: $r/H \approx 0.3,~r/\eta
\approx  
18,670$; $\diamond$: $r/H
\approx 3,~r/\eta \approx 1,86,700$. ---, Gaussian.
Figure (a) is  
for the unfiltered signal; (b) is for data high-pass filtered at $f_o = 0.025$ Hz, $r_f/H \approx 9$, $r_f/\star \approx 120,000$; (c) $f_o$ =
2 Hz, $r_f/H \approx 0.11,~r_f/\eta \approx 1,500$; (d) $f_o$ = 25 Hz, $r_f/H
\approx 8.8 \times 10^{-3},~r_f/\eta \approx 116$. 
}}
\label{fig:pdf_vel_incr_filter_unfilter}
\end{center}
\end{figure}

\newpage
\begin{figure}[ht]
\begin{center}
\parbox{5.5in}{
\psfig{file=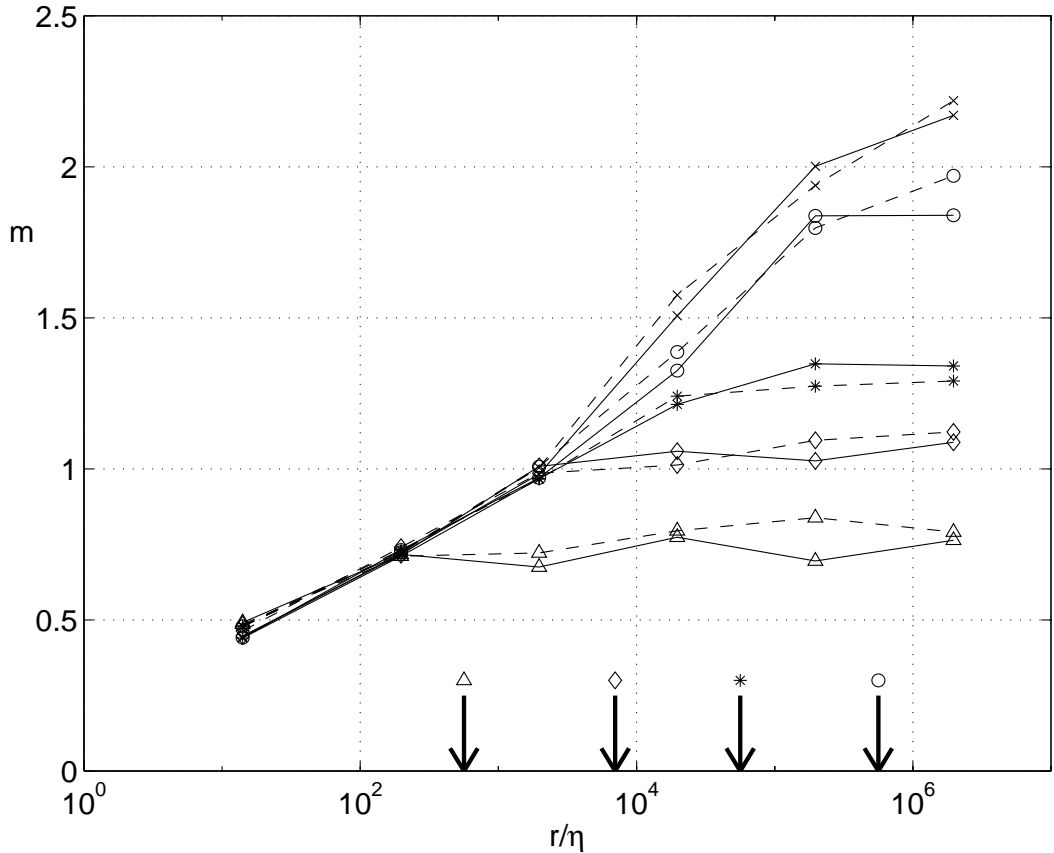,width=5.5in,rwidth=5.5in}
}
\caption{\small{The stretching exponent, $m$, versus the separation
distance for  
the unfiltered and filtered data sets.  The values of $m$ are obtained by
fitting empirically
$p_{\Delta u_r} = p(0)exp[-c|\Delta u_r|^{m(r)}]$ over the entire range of
$\Delta  
u_r$ except towards the tails where there is some scatter because of
unconverged  
statistics. --- is for the positive part of the pdf, and $-~-$ is for the negative part. $\times$, unfiltered; $\circ$, high-pass filtered at $f_o = 0.025$ Hz ($r_f/H \approx 8.8$); $\ast$, $f_o$ 0.25 Hz ($r_f/H \approx 0.88$); $\diamond$, $f_o$ = 2 Hz ($
r_f/H \approx 0.11$); $\triangle$, $f_o$ = 25 Hz ($r_f/H \approx 0.009$).  Arrows mark the corresponding $r_f/\eta$ values. For the unfiltered data, the pdf tends essentially to a Gaussian ($m \approx 2$)---actually slightly sub-Gaussian because $m>2$; fo
r the filtered cases, the asymptotic shapes depend on the respective cut-off scales.}}
\label{fig:str_exp_fit}
\end{center}
\end{figure}

\newpage
\begin{figure}[ht]
\begin{center}
\parbox{5.5in}{
\psfig{file=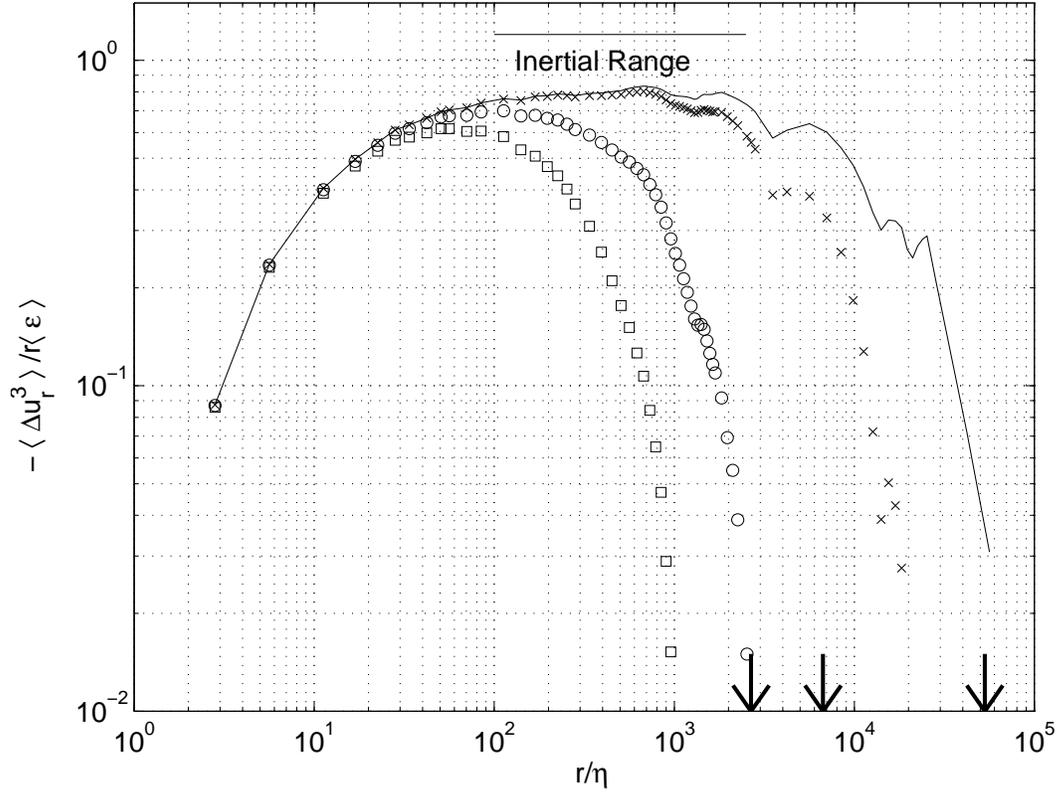,width=5.5in,rwidth=5.5in}
}
\caption{\small{The Kolmogorov function $\la \Delta u_r^3 \ra/r \la
\varepsilon \ra$  
plotted for unfiltered (---) and high-pass filtered traces ($\times$, $f_o$ =
0.25  
Hz, $r_f/H \approx 0.88$; $\circ$, $f_o$ = 2 Hz, $r_f/H \approx 0.11$; $\Box$, $f_o$ = 5 Hz, $r_f/H \approx 0.044$). The normalizing scales
$\eta$ and  
$\la \varepsilon \ra$ are for the unfiltered signal, but the filtering does
not  
change them to any significant extent. The horizontal line extending between  
$r/\eta$ of 100 and 2500 is the estimated inertial range. In this figure
and others  
to follow, the cut-off scale $r_f$ is shown by an arrow on the abscissae. The removal of large scales affects
scales that are typically two decades smaller than $r_f$.}}
\label{fig:K_func_filter_unfilter}
\end{center}
\end{figure}

\newpage
\begin{figure}[ht]
\begin{center}
\parbox{5.5in}{
\psfig{file=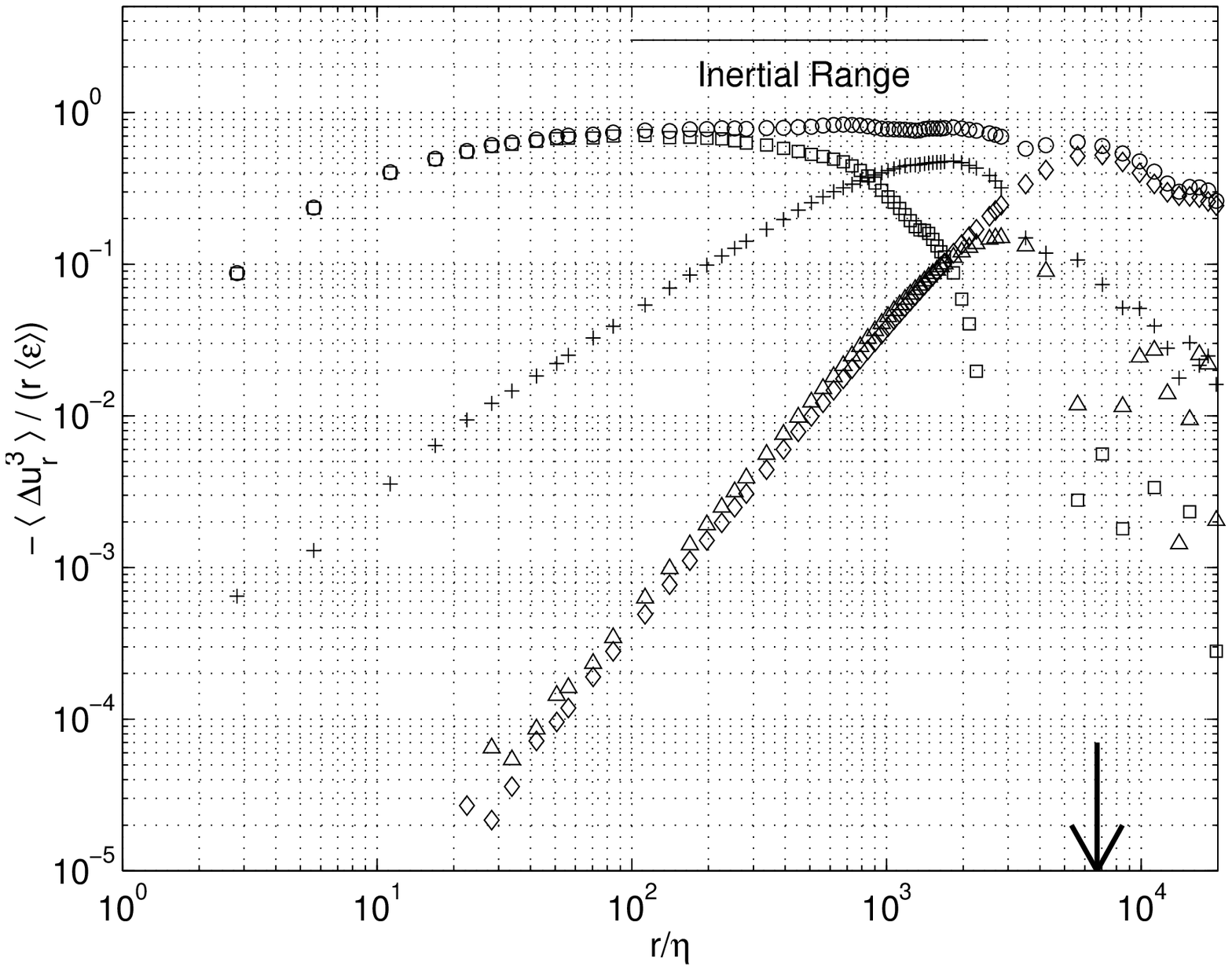,width=5.5in,rwidth=5.5in}
}
\caption{\small{The result of decomposing the third-order structure function $\la \Delta u_r^3 \ra$
in terms of  $\Delta u_r^<$ and $\Delta u_r^>$; $f_o = 2$ Hz, $r_f/H \approx$ 0.11. $\circ$, $\la \Delta u_r^3 \ra/r \la \varepsilon \ra$; $\Box$, $\la (\Delta u_r^>)^3 \ra/r \la \varepsilon \ra$; $+$, 3$\la (\Delta u_r^>)^2 \Delta u_r^< \ra/r \la \varepsilon \ra$; $\triangle$, 3$\la (\Delta u_r^<)^2 \Delta u_r^> \ra/r \la \varepsilon \ra$; $\diamond$, $\la (\Delta u_r^<)^3 \ra/r \la \varepsilon \ra$. In the upper part of the inertial range, the most dominant term is $3 \la (\Delta u_r^>)^2 \Delta u_r^< \
ra$.}}
\label{fig:third_order_decomposed}
\end{center}
\end{figure}

\newpage
\begin{figure}[ht]
\begin{center}
\parbox{5.5in}{
\psfig{file=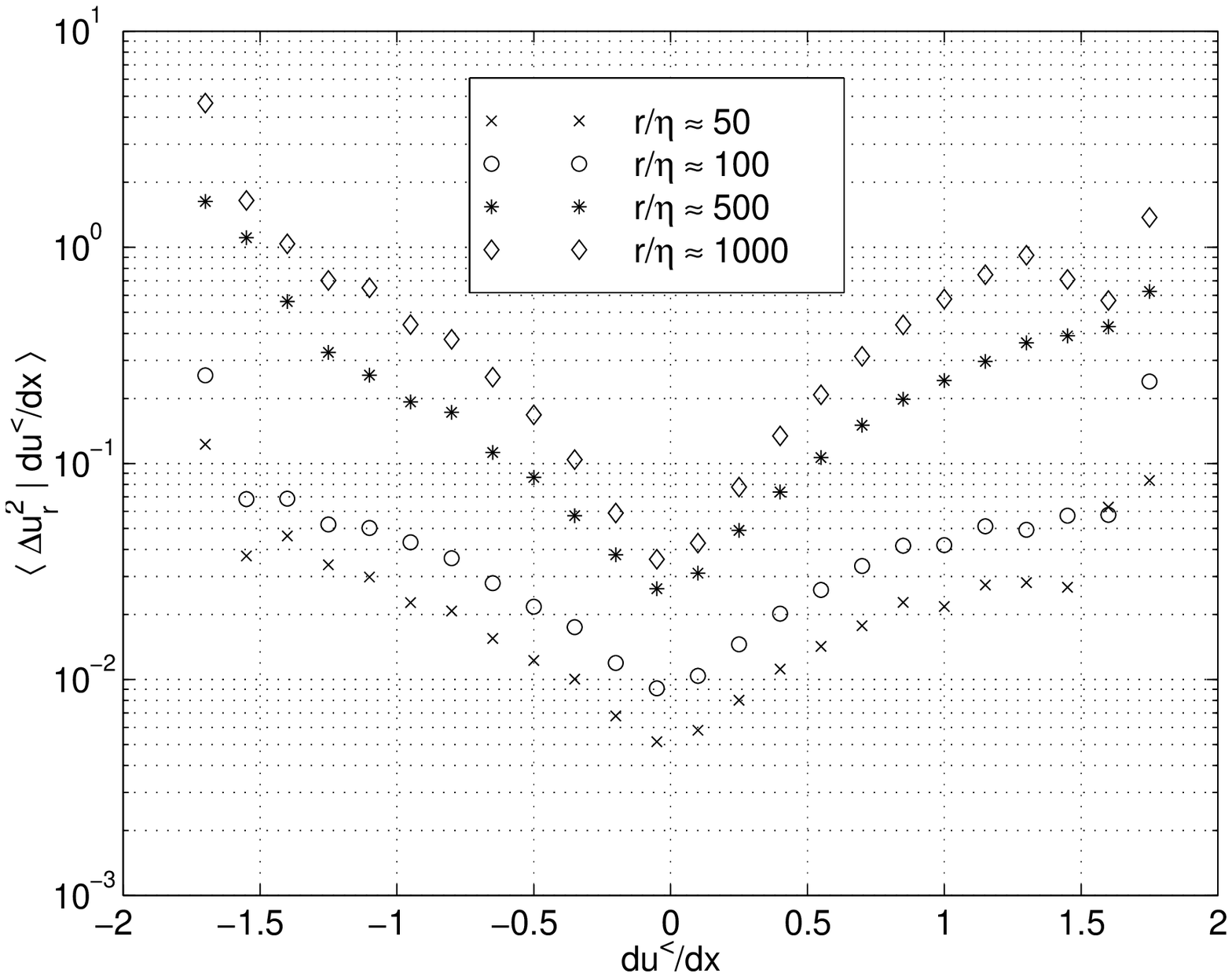,width=5.5in,rwidth=5.5in}
}
\caption{\small{The conditional expectation of the unfiltered variance $(\Delta  
u_r)^2$ conditioned on the rate of strain of the largest scale retained
in the  
filtering operation, $d\Delta u_r^</dx$; $f_o = 2$ Hz, $r_f/\eta \approx$ 7,000. The various symbols correspond to different scale separations $r$, as noted in the inset.}}
\label{fig:cond_stfc_on_dudx}
\end{center}
\end{figure}

\newpage
\begin{figure}[ht]
\begin{center}
\parbox{5.5in}{
\psfig{file=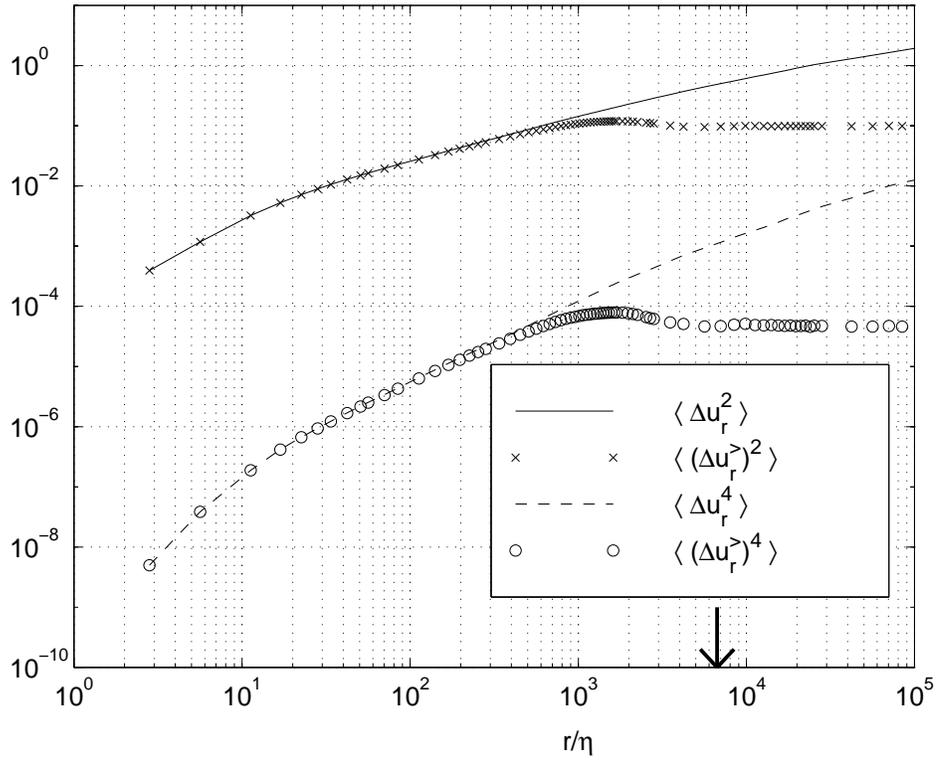,width=5.5in,rwidth=5.5in}
}
\caption{\small{ Second and fourth order structure functions for the
unfiltered (lines) and filtered data (symbols). ---  and $\times$ are for the second-order; $-~-$ and $\circ$ are for the fourth-order; $f_o = 2$ Hz, $r_f/\eta \approx$ 7,000.}}
\label{fig:sec_fourth_filter_unfilter}
\end{center}
\end{figure}

\newpage
\begin{figure}[ht]
\begin{center}
\parbox{5.5in}{
\psfig{file=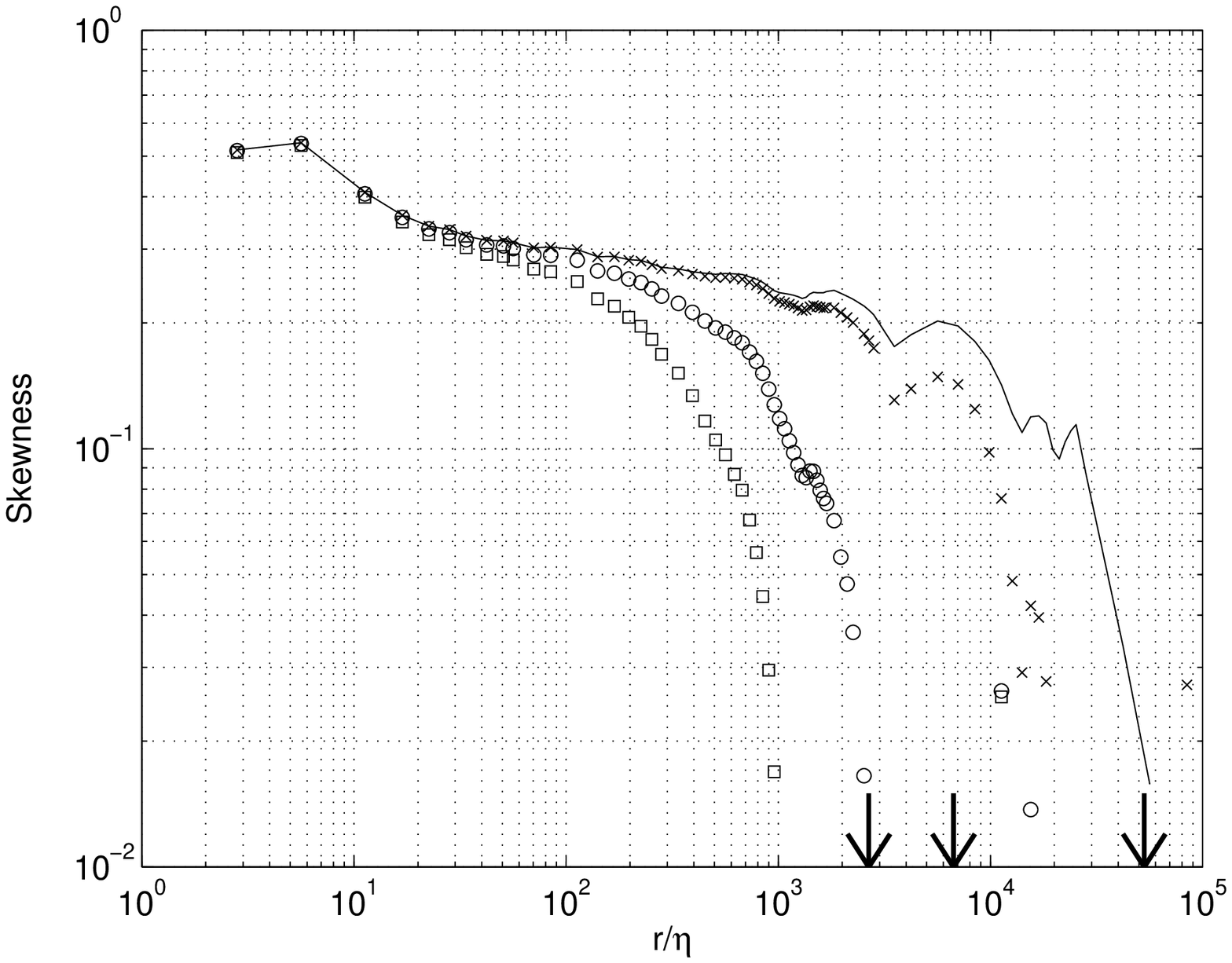,width=5.5in,rwidth=5.5in}
}
\caption{\small{The skewness of velocity increments for the unfiltered (---) and
filtered  
signals ($\times, f_o = 0.25$ Hz, $r_f/H \approx 0.88$; $\circ, f_o = 2$ Hz, $r_f/H \approx$ 0.11; $\Box$, $f_o = 5$ Hz, $r_f/H \approx$ 0.044). The filter settings are shown by arrows. 
}}
\label{fig:skewness_filter_unfilter}
\end{center}
\end{figure}
\newpage
\begin{figure}[ht]
\begin{center}
\parbox{5.5in}{
\psfig{file=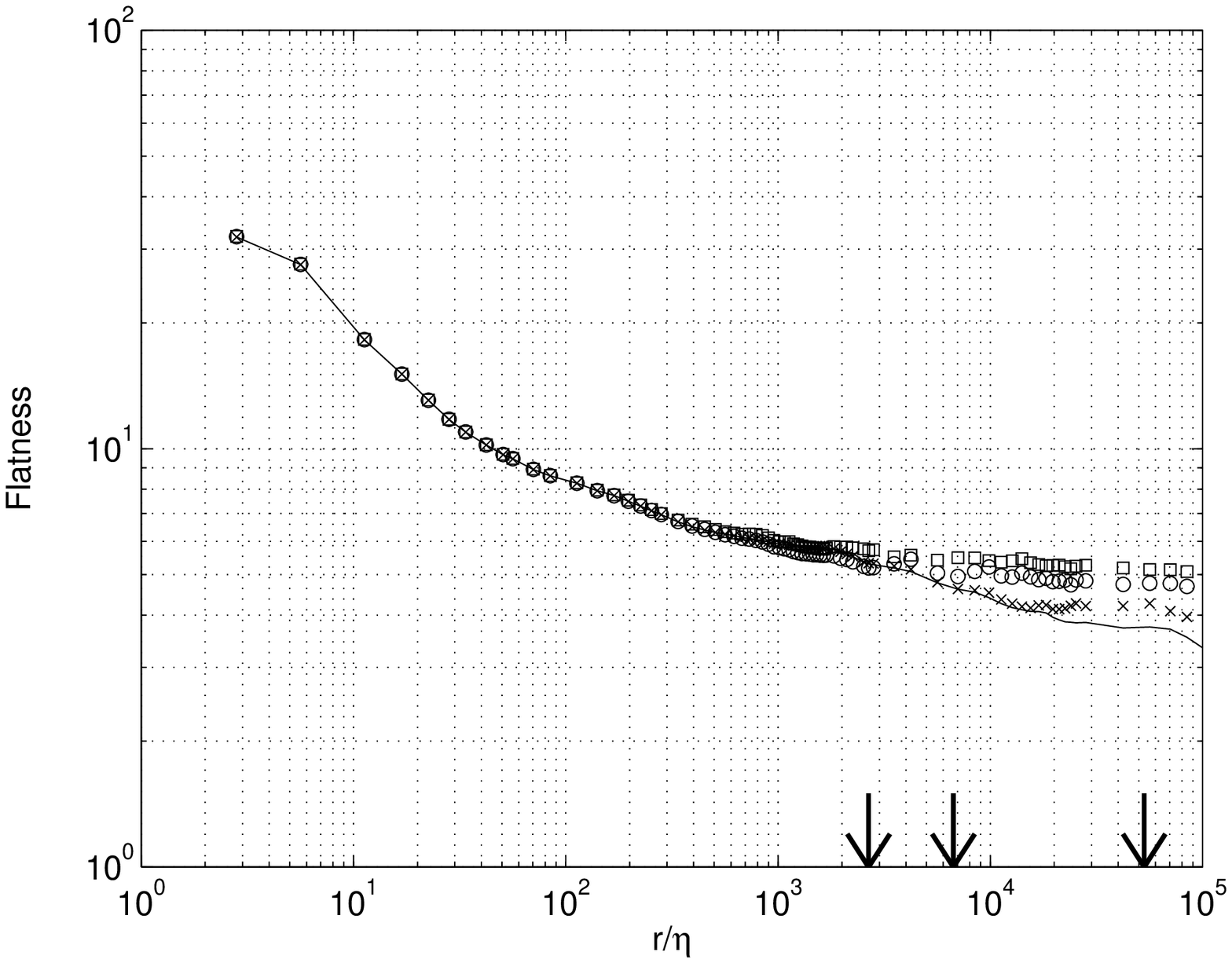,width=5.5in,rwidth=5.5in}
}
\caption{\small{The flatness factor of velocity increments for the unfiltered (---) and  
filtered signals ($\times, f_o = 0.25$ Hz, $r_f/H \approx 0.88$; $\circ, f_o = 2$ Hz, $r_f/H \approx$ 0.11; $\Box$, $f_o = 5$ Hz, $r_f/H \approx$ 0.044). The filter settings are shown by arrows.}}
\label{fig:flatness_filter_unfilter}
\end{center}
\end{figure}

\newpage
\begin{figure}[ht]
\begin{center}
\parbox{5.5in}{
\psfig{file=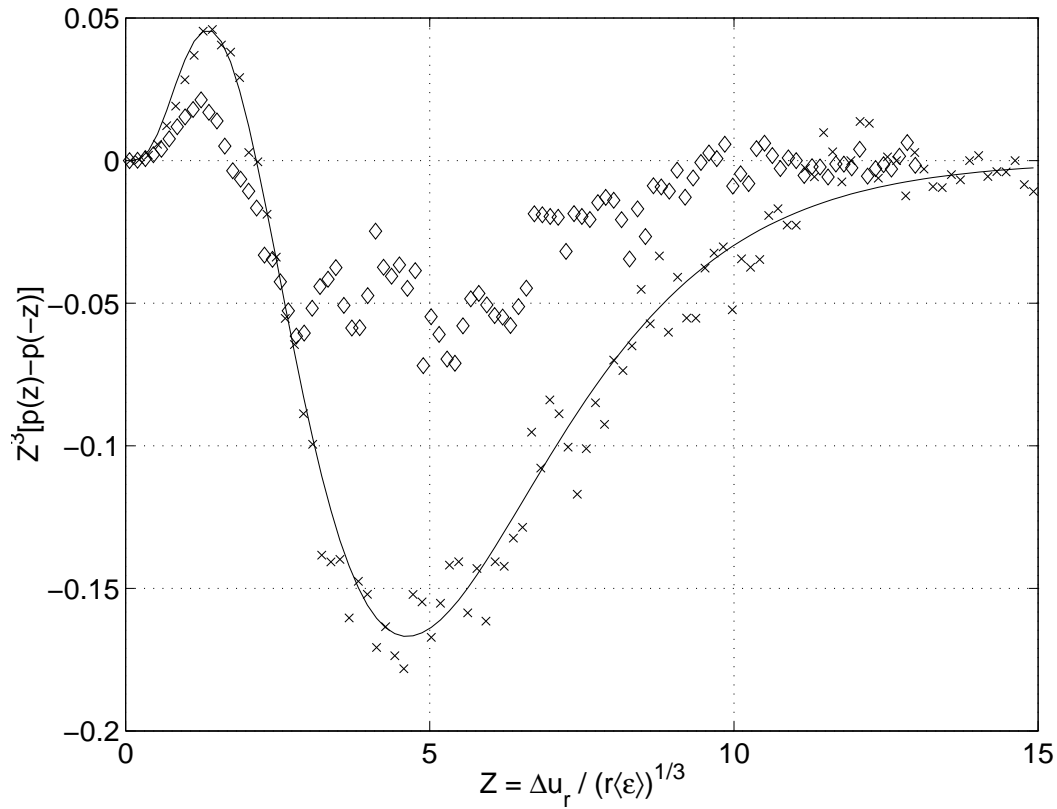,width=5.5in,rwidth=5.5in}
}
\caption{\small{The integrand of $K \equiv \la \Delta u_r \ra^3/r \la \varepsilon \ra$, plotted against $\Delta u_r/(r \la \varepsilon \ra)^{1/3}$. The crosses are for the unfiltered signal, $r/\eta = 1000$, and --- represents a smooth line through them. 
The corresponding quantity for the filtered data, with $f_o= 2$ Hz, $r_f/\eta \approx$ 7000, are shown by diamonds. The effect of the filter is drastic at all scales. The relatively large scatter in the data is the result of taking the difference between 
two nearly equal quantities.}}
\label{fig:skew_integrand}
\end{center}
\end{figure}

\end{document}